\documentclass[final,authoryear,5p,twocolumn,10pt]{elsarticle}

\usepackage{graphics}
\usepackage{graphicx}
\usepackage{epsfig}

\usepackage{amssymb}
\usepackage{amsthm}
\usepackage{amsmath}

\usepackage{lineno}

\newcommand{\be}{\begin{equation}}
\newcommand{\ee}[1]{\label{#1}\end{equation}}
\newcommand{\re}[1]{Eq. (\ref{#1})}
\newcommand{\rr}[2]{Eqs (\ref{#1}) and (\ref{#2})}
\newcommand{\pa}[2]{\frac{\partial{#1}}{\partial{#2}}}
\newcommand{\od}[2]{\frac{{ d}{#1}}{{ d}{#2}}}

\newcommand{\R}[1]{Equation (\ref{#1})}

\newcommand{\bn}{\begin{eqnarray}}
\newcommand{\en}{\end{eqnarray}}
\newcommand{\ba}{\begin{array}}
\newcommand{\ea}{\end{array}}

\newcommand{\inl}{\int\limits}

\newcommand{\sml}{\sum\limits}

\newcommand{\bit}{\begin{itemize}}
\newcommand{\eit}{\end{itemize}}
\newcommand{\bce}{\begin{center}}
\newcommand{\ece}{\end{center}}
\newcommand{\D}{\displaystyle}

\newcommand{\sub}[1]{\scriptscriptstyle{#1}}

\journal{arXiv.org}

\begin{document}
%\begin{linenumbers}
\begin{frontmatter}



\title{\textbf{Absorption of Soluble Gases by Atmospheric Nanoaerosols}}


\author[1]{\large{Tov Elperin}}
\author[1]{\large{Andrew Fominykh}}
\author[1]{\large{Boris Krasovitov}\corref{cor1}}
\author[2]{\large{Alexey Lushnikov}}
\address[1]{\normalsize{Department of Mechanical Engineering, The Pearlstone Center for Aeronautical Engineering Studies,Ben-Gurion University of the Negev, P.O.Box 653, 84105 Israel}}
\address[2]{\normalsize{Karpov Institute of Physical Chemistry, 10, Vorontsovo Pole, 105064 Moscow Russia}}
\cortext[cor1]{Corresponding author: borisk@bgu.ac.il}
\begin{abstract}
We investigate mass transfer during absorption of atmospheric trace soluble gases by a single droplet whose size is comparable to the molecular mean free path in air at normal
conditions. It is assumed that the trace reactant diffuses to the droplet surface and then reacts with the substances inside the droplet according to the first order rate law. Our analysis  applies a flux-matching theory of transport processes in gases and assumes constant thermophysical properties of the gases and liquids. We derive an integral equation of Volterra type for the transient molecular flux density to a liquid droplet and solve it numerically. Numerical calculations are performed for absorption of sulfur dioxide ($\mathrm {SO}_2$), dinitrogen trioxide (N$_{2}$O$_{3}$) and chlorine (Cl$_{2}$) by liquid nanoaerosols accompanied by chemical dissociation reaction. It is shown that during gas absorption by nanoaerosols the kinetic effects play significant role, and neglecting kinetic effects leads to significant overestimation of the soluble gas flux into a droplet during all the period of gas absorption.
\end{abstract}

\begin{keyword}
nanoaerosol, gas absorption, diffusion, free molecular flow, soluble gas, chemical dissociation reaction

\end{keyword}

\end{frontmatter}

% \linenumbers

\section{Introduction}
%\label{1}
%Wet removal of gaseous pollutants by cloud droplets is involved in
%various atmospheric processes such as scavenging of gaseous pollutants, acid deposition etc. Heat and mass transfer during gas absorption by liquid droplets is important in various fields of environmental engineering and atmospheric science. Clouds represent an important element in self-cleansing process of the atmosphere \citep{Flossmann1998}.
Atmospheric aerosols are directly emitted into the atmosphere from natural or anthropogenic sources or can be formed in the atmosphere through nucleation of gas-phase species. Aerosol nucleation events produce a large fraction of atmospheric aerosols. New particle formation occurs in two distinct stages, i.e., nucleation to form a critical nucleus and subsequent growth of the critical nucleus to a larger size ($>$2 -- 3 nm) that competes with capture and removal of the freshly nucleated nanoparticles by coagulation with pre-existing aerosols \citep{Zhang2012}. In the continental boundary layer, there are frequent observations of the formation of ultrafine aerosol particles accompanied by the subsequent growth \citep{Kulmala2004}. Gas absorption of soluble trace atmospheric gases by the liquid atmospheric aerosol particles including ultrafine particles plays an important role in climate and atmospheric chemistry.

The consequence for the aerosol climate forcing is that the cooling can be intensified with increasing atmospheric amount of water-soluble trace gases such as $\mathrm{HNO}_3$, counteracting the warming effect of the greenhouse gases \citep{Kramer2000}.
Scavenging of atmospheric gaseous pollutants by cloud droplets is a result of gas absorption mechanism \citep{Pruppacher1997,Flossmann1998}. Gas scavenging of highly soluble gases by atmospheric water droplets includes absorption of $\mathrm{HNO}_3, \mathrm H_2 \mathrm O_2, \mathrm H_2 \mathrm{SO}_4, \mathrm{HCL}$ and some other
 gases. The sources of these gases in the atmosphere are briefly reviewed in \citet{Seinfeld2006} and \citet{Hayden2008}. Soluble gas absorption by noncirculating droplets was investigated experimentally by \citet{Taniguchi1992} where conditions of noncirculation for falling liquid droplets were determined by employing water droplets with Sauter mean diameter equal to 0.185, 0.148 and 0.137 mm.

 Gas absorption by stagnant liquid droplets in the presence of inert admixtures when both phases affect mass transfer was analyzed by \citet{Clift1978}, pp. 54-55. Scavenging of soluble gases by single evaporating droplets was studied by \citet{Elperin2007,Elperin2008}.

All aforementioned works considered the trace gas absorption in the continuous limit, where Fick's law relates the flux and the concentration gradient of reactant. Recently some studies attempt to describe the reactant transport in the gaseous phase in the
free--molecular and the transition regimes, where the droplet size is less than or comparable to the mean free path of the molecules in the gaseous phase. Comprehensive reviews of the results and approaches can be found in \citet{Seinfeld2006}, \citet{Clement2007}, \citet{Poschl2007} and \citet{Lushnikov2012}. The existing attempts to describe the transitional and free--molecule regimes encounter some difficulties in formulating the boundary conditions to the transport equations. For example, \citet{Vesala2001} used the diffusion equations and a microscopic boundary condition for describing the reactant  transport toward the particle surface. The rigorous approach requires solution of the full transport problem including the reactant transport in the gaseous phase and gas--liquid interface by solving the Boltzmann kinetic equation and the diffusion--reaction equation inside the droplet. To the best of our knowledge this approach was not applied as yet.

In this study we employ a different approach for describing gas uptake by nanodroplets. We modify the flux--matching approach of \citet{Lushnikov2004} by including the in--particle chemical transformations of the reactant molecules. The results are applied for considering the trace gases scavenging by atmospheric aerosols.

Let us assume that the reactant molecules ($A$--molecules) move toward the droplet which captures them. The further fate of reactant molecules depends on the results of chemical reactions inside the particle. Let us denote by $n_\pm$ the concentrations of $A$--molecules right  outside  ($n_+$) or right inside ($n_-$) the particle surface. Clearly these concentrations $n_\pm$ depend on the nature of physicochemical processes at the surface and inside the particle. Let $n_\infty$ be the number density of $A$--molecules far from the particle. It is commonly accepted that the concentration difference $n_\infty-n_+$ drives a flux of $A$--molecules towards the particle surface. The mass of the particle begins to increase and its chemical composition changes. The rate of change of the number of $A$--molecules inside the particle is equal to the total flux $J$, i.e. the total number of molecules deposited per unit time at the particle surface minus the rate of consumption of molecules $A$ by chemical reaction inside the particle. Some fraction of $A$--molecules is assumed to escape from the particles. In steady--state conditions the flux $J$ can be written as
\begin{equation}\label{2}
    J=\alpha(a)(n_\infty-n_+)\:,
\end{equation}
where $\alpha(a)$ is the capture efficiency and $a$ is the particle radius. Clearly, capture efficiency $\alpha$ depends on  the mass accommodation coefficient $S_p$. The latter is defined as the probability for an $A$--molecule to stick to the particle after a single collision. Since the interface and in--particle  processes determine the value of $n_+$ Eq. (\ref{2}) can be rewritten as follows \citep{Lushnikov2012}:
\begin{equation}\label{3}
J=\frac{\alpha(a)n_\infty}{1+\alpha(a)\psi(a,J)}\:,
\end{equation}
where $\psi(a,J)$ is a function depending on the nature of the chemical reaction. In the case of the first--order chemical reaction the function $\psi$ is independent of $J$. An example of such function is given below. If, however, the chemical reaction inside the
particle is of the higher order, then $J$  is a solution of the transcendent  equation  (\ref{3}). Note that we did not yet specify the functions $\alpha(a)$, $\psi(a,J)$, and \re{3} contains only $n_\infty$. The information on the processes at the surface and inside the particle is carried by the function $\psi(a,J)$.  Hence, \re{3} is quite general. All further approximations  concern the values of the uptake efficiency $\alpha(a)$ and the reaction function $\psi(a)$.

\section{Description of the model}
%\label{s2}
\subsection{Preliminary remarks}
The characteristic times of changes of the number density of  reactant $A$ in  gaseous and liquid phases differ by several orders of magnitude. In particular, the relaxation time inside the micrometer-sized droplet is $\tau_{\sub{L}}\propto a^2/D_{\sub{L}}\simeq 10^{-3}$ s, where $D_{\sub{L}}\simeq 10^{-5}$ cm$^2$/s  denotes the diffusivity  of the reactant molecules in the liquid  phase. The relaxation time in the gaseous phase can be estimated as $\tau_{\sub{G}}\propto a/v_{\sub{T}}\simeq 10^{-8}$ s, where $v_{\sub{T}}$ is the mean thermal velocity of the reactant molecules, $v_{\sub{T}}\simeq 10^2$ m/s. Here one can use the free molecular estimate because the droplet size is of the order of the mean free path of the reactant molecule. Following \citet{Seinfeld2006} the characteristic time of diffusion in a gaseous phase, corresponding to the time required by gas-phase diffusion to establish a steady-state profile around a particle, can be alternatively estimated as $\tau_{\sub{G}}\propto a^2/D_{\sub{G}}\simeq 10^{-8}$ s.  As can be seen from these estimates, the characteristic time of diffusion in a gaseous phase, $\tau_{\sub{G}}$ , is much smaller than the characteristic time of diffusion in the liquid phase, $\tau_{\sub{L}}$, which is required for a saturation of the droplet by soluble gas (i.e., $\tau_{\sub{G}} \ll \tau_{\sub{L}}$). Therefore for the large values of $t$ ($t \gg \tau_{\sub G}$) it is reasonable to assume that concentration profile in a gaseous phase in the transitional regime and the flux attain their quasi-steady state values \citep{Wagner1982} and are determined by Eq. (\ref{2}) (or in more general form by Eq. (\ref{3})).

The molecular mean free path in air at normal conditions is $\ell\approx$ 65 nm , i.e., it is comparable to the sizes of the nanometer droplets. This implies that the motion of the reactant molecules cannot be described as the Fickian diffusion, and one must
apply the Boltzmann kinetic theory. However, solving the Boltzmann equation analytically or numerically is a formidable task.

The idea to apply the flux--matching approach in the aerosol kinetics was pioneered by \citet{Fuchs1964}. His reasoning was quite simple. At large distance from the droplet the reactant transport can be described by the diffusion equation. In the vicinity of the droplet at the distances of the order of $\ell$ or less the collisions with the carrier gas do not hinder the reactant transport. Consequently, inside the region $a<r< R\propto\ell$ ($R$ is referred to as the radius of the limiting sphere) the reactant molecules move in the free molecule regime. The value of $R$ must be found from different consideration. \citet{Fuchs1971} proposed to determine this value from the numerical solution of the BGK equation \citep[see e.g.,][]{Sahni1966}. An improved version of the Fuchs interpolation formula \citep{Fuchs1964} was obtained by \citet{Loyalka1982} in near continuum regime by solving Boltzmann equation using momentum method.

\subsection{Trapping efficiency}

The latest modification of the Fuchs theory \citep{Lushnikov2004,Lushnikov2012} includes the solution of the diffusion equation with a fixed  flux $J$ in the diffusion zone $r>R$, the solution of the collisionless Boltzmann equation in the free molecular zone $r<R$, and determining the radius of the limiting sphere from the condition of equality of the fluxes in both zones. The expression for $\alpha(a)$ was found by \citet{Lushnikov2004} for $n_+=0$ and $S_p=1$:
\begin{equation}\label{4}
\alpha(a)=\frac{2\pi a^2v_{\sub{T}}}{1+\D\sqrt{1+\left(\frac{av_{\sub{T}}}{2D_{\sub{G}}} \right)^2}}\; ,
\end{equation}
where $D_{\sub G}$ is the reactant diffusivity in the carrier gas. The extension of this formula to the case $n_+\ge0$ and $S_p\le1$ reads \citep[for details see][]{Lushnikov2012}:
\begin{equation}\label{5}
\alpha(a)=\frac{S_p\pi
a^2v_T}{1+\D\frac{S_p}2\left[\D{\sqrt{1+\left(\frac{av_{\sub T}}{2D_{\sub G}}\right)^2}}-1\right]}\; .
\end{equation}

The radius $R$ of the limiting sphere is found from the condition of the equality of flux in the diffusion region to the flux from in free molecular region. This condition yield the radius of the limiting sphere \citep{Lushnikov2012,Lushnikov2004}:
\begin{equation}\label{7}
R=\sqrt{a^2+\left(\frac{2D_{\sub G}}{v_{\sub T}}\right)^2}.
\end{equation}

It must be noted that $R$ is independent of $S_p$ and $n_+$. The spherical surface with the radius $R$ separates between the zones of the free--molecular and the continuous flow regimes. The value of $2D_{\sub G}/v_{\sub T}$ is of the order of $\ell$, the reactant molecule mean free path. Hence, if $a\simeq\ell$ or less, then the radius $R$ is of the order of $R\simeq\ell$.

The concentration profile of the reactant $n(r)$ in the gaseous phase inside the limiting sphere $a<r<R$ is continuous at $r=R$ together with its first derivative and is given by the following formula \citep{Lushnikov2004}:
\begin{equation}\label{8}
\frac{n(r)-n_+}{n_\infty-n_+}=\left(1-\frac{\alpha(a)}{4\pi D_{\sub G}R}\right)\frac{b(r)}{b(R)}
\end{equation}
where
\begin{equation}\label{8a}
b(r)= 1-\frac{S_p}2\left(1-\sqrt{1-\frac{a^2}{r^2}}\right).
\end{equation}
Outside the limiting sphere at $r\geq R$
\begin{equation}\label{9}
\frac{n(r)-n_+}{n_\infty-n_+}=1-\frac{\alpha(a)}{4\pi Dr}\:.
\end{equation}
Note that the number density $n(a)$ is always larger than $n_+$. The formula for the concentration jump at the particle surface reads \citep{Lushnikov2012}:
\begin{equation}\label{10}
\Delta_a=n(a)-n_+=(n_\infty-n_+)\left(1-\frac{\alpha(a)}{4\pi D_{\sub G}R} \right)\frac{b(a)}{b(R)}.
\end{equation}
Inspection of \re{5} shows that when $a v_{\sub T} / 2 D_{\sub G} \gg 1$, $\alpha (a) = 4 \pi a D_{\sub G}$ and \re{5} recovers the Maxwell's equation for the molecular flux in the continuum regime $J_c$:
\begin{equation}\label{maxwell}
J_c=4\pi a D_{\sub G} (n_\infty-n_+).
\end{equation}

\begin{figure}[h!]
 \begin{center}
  \includegraphics[width= 8 cm]{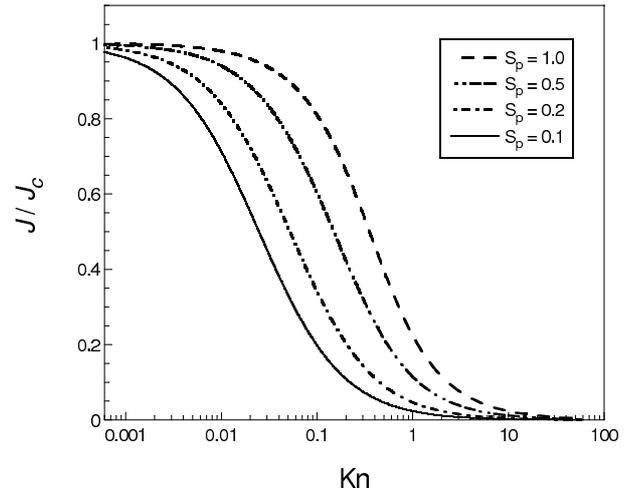}
  \caption[Figure 1]{Ratio of the molecular flux to the molecular flux in a continuum regime, $J/J_c\,$, as a function of Knudsen number $Kn$.}\label{fig1}
  \end {center}
\end{figure}

\begin{figure}[h!]
 \begin{center}
  \includegraphics[width= 8 cm]{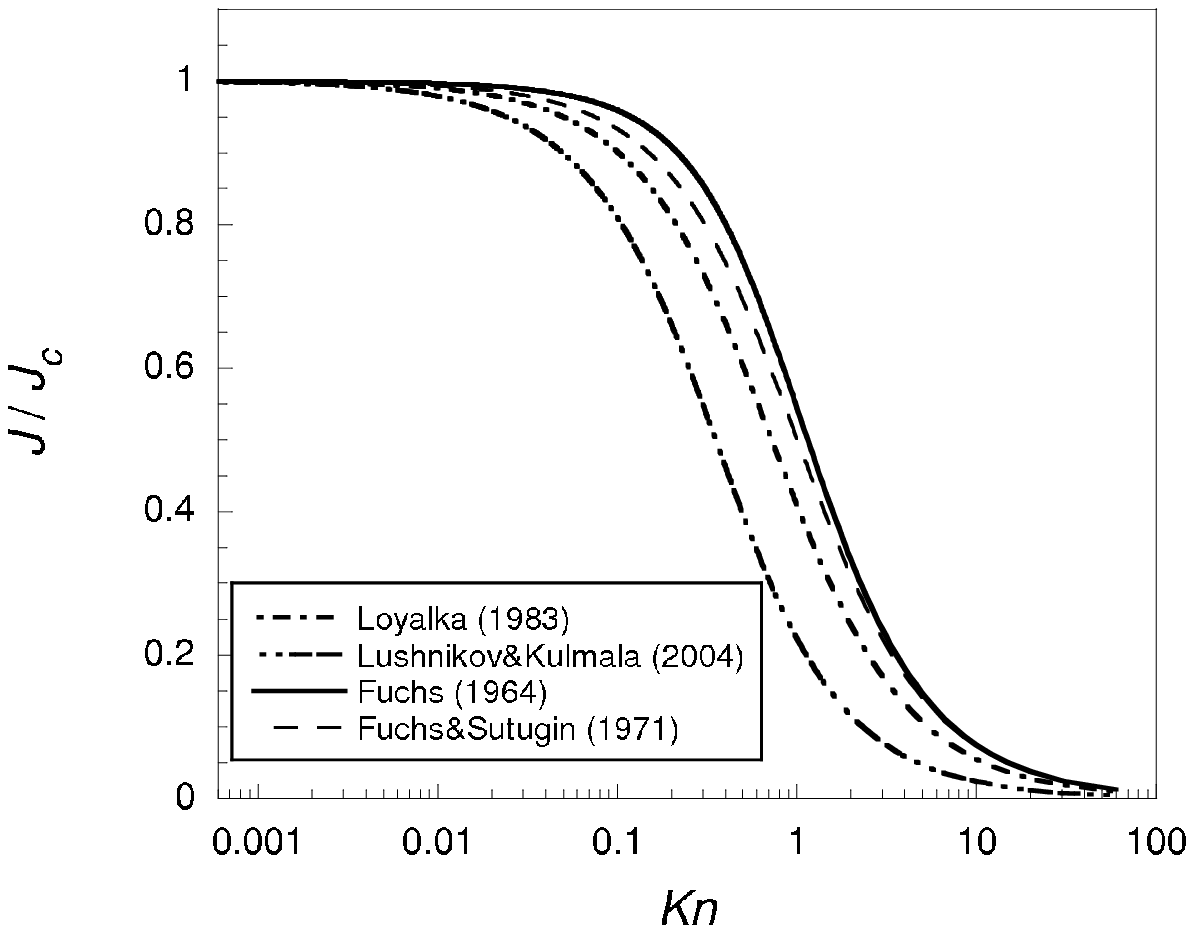}
  \caption[Figure 1b]{Ratio of the molecular flux to the molecular flux in a continuum regime, $J/J_c\,$, as a function of Knudsen number $Kn$: \textbf{------} - \citet{Fuchs1964}; \textbf{-- -- --} - \citet{Fuchs1971}; \textbf{-- $\cdot$ --} - \citet{Loyalka1982}; \textbf{-- $\cdot\cdot$ --} - \citet{Lushnikov2004}. Accomodation coefficient $S_p = 1$.}\label{fig1b}
  \end {center}
\end{figure}

In Fig. \ref{fig1} we showed the dependence of the ratio $J/J_c$ vs. Knudsen number $Kn$ for differen values of the accommodation coefficient $S_p$ ($S_p$ was assumed to be 0.1, 0.2, 0.5 and 1.0 and $D_{\sub G}\simeq 10^{-5}$ m$^2$/s). As can be seen from this plot the role of the kinetic effects can be significant for the $Kn \gtrsim 0.1$.
Comparison of mass transfer rates as function of $Kn$ predicted by different theories \citep{Fuchs1964, Fuchs1971, Loyalka1982, Lushnikov2004} are shown in Fig. \ref{fig1b}. As can be seen from Fig. \ref{fig1b} all approaches yield approximately the same results for small $Kn$ numbers, $Kn\lesssim 0.1$, and for $Kn\gtrsim 10$.

\subsection{Inner diffusion--reaction equation}

Consider now the effect of the first order chemical reaction, e.g., chemical reaction dissociation, inside the droplet on the reactant flux towards the droplet. Neglecting recombination the number density of reactant molecules inside the particle $n_L(r,t)$ is governed by the linear diffusion--reaction equation:
\be
\pa{n_L}{t}=D_L\Delta n_L - \lambda n_L,
\ee{1a}
where $n_L=n_L(r,t)$ is the number density of reactant molecules inside the particle, $D_L$ is the reactant diffusivity in the liquid phase, and $\lambda$ is the dissociation rate.
\R {1a} must be supplemented with the initial  condition,
\be
n_L(r, 0)=0
\ee{2a}
(no reactant inside the droplet at $t=0$)and the boundary conditions:
\begin{equation}\label{bc1}
    \left.\pa{n_L(r,t)}{r}\right|_{r=0} = 0
\end{equation}
and
\be
j(t)=-D_L\left.\pa{n_L(r,t)}{r}\right|_{r=a},
\ee{3a}
where $j=j(t)$ is the flux density $j=J/(4\pi a^2)$, $J$ is the total flux.

Equation (\ref{1a}) with the initial and boundary conditions (\ref{2a}) -- (\ref{3a}) can be solved by the method of separation of variables (see Appendix A). The result is the concentration profile $n_L(r,t)$ as a linear functional of $j(t)$. Substituting the determined concentration distribution $n_L(r,t)$
into the boundary condition, \re{3a}, yields the integral equation of Volterra type \citep{Apelblat2008} for $j(t)$:
\be
j(t)=\frac{\alpha(a)}{4\pi a^2}\left[n_\infty-{\cal H}\inl_0^tS(t-t')j(t')dt'\right],
\ee{j}
where
\be
S(\xi)=2\sml_{n >\;0} e^{-[D_L(\mu_n / a)^2+\lambda]\xi}+3 e^{-\lambda\xi}
\ee{S}
and $\mu_n$ is the infinite set of the roots of the following transcendental equation:
\begin{equation}\label{teq}
    \mu = \tan(\mu)\:.
\end{equation}
\R {j} can be rewritten in the dimensionless form for the dimensionless flux $\displaystyle j^\ast (\tau)= j(t)4\pi a^2/\alpha(a)n_\infty \;$, $\tau = D_{\sub L}t/a^2$:
\be
j^\ast(\tau)=1 - g(a)\inl_0^{\tau}S^*(\tau-\tau')j^*(\tau')d\tau',
\ee{j1}
where $\D g (a)= \frac{3\alpha(a)\cal H}{4\pi a D_L}$ and
\be
S^*(\tau - \tau')= e^{-\mathrm {D}a(\tau - \tau')}\left[1 + \frac{2}{3}\sum_{n=1}^\infty e^{-\mu_n^2(\tau - \tau')}\right]\:.
\ee{K}
In Eqs. (\ref {j1}) and (\ref {K}) $D\mathrm a = \lambda a^2 / D_L$ is Damkohler number, $\cal H$ $= (H_A \mathcal{R} \mathnormal {T})^{-1}$ and $H_A$ is Henry's law constant, $\mathcal {R}$ is universal gas constant and $T$ is the temperature in the gaseous phase.

\section{Numerical method}

For the solution of the integral equation \re{j1} we use the method based on the approximation of the integral in \re {j1} using some quadrature formula:
\begin{equation}\label{n1}
\int_a^b F\left(x\right)dx =\sum_{j=1}^n A_j F\left(x_j\right)dx + R_n\left(F\right),
\end{equation}
where $x_j\in[a,b]$, $j=1,2,...,n$, $A_j$ are the coefficients
associated with a family of quadrature rules and
$R_n\left(F\right)$ is a corresponding residuum. Taking
successively $x = x_i$ $(i=1,...,n)$ and using the quadrature formula
after discarding the terms $R_n\left(F_i\right)$ $(i=1,...,n)$ we obtain the following system of linear algebraic equations:

\begin{equation}\label{n2}
    j^*_i - g(a)\sum_{j=1}^n A_jS^*_{ij}j^*_j = 1\;\;\;\;\; (i=1,...,n).
\end{equation}

The solution of Eqs. (\ref{n2}) yields the approximative value of the
unknown function $j_{i}$ at the mesh point $\tau_{i}$. The system
of Eqs. (\ref{n2}) can be written in the following form:

\begin{equation}\label{n3}
    -\sum_{j=1}^{i-1}A_j K_{ij} j^*_j +\left(1-A_i K_{ii}\right) j^*_i = 1,
\end{equation}
where $K_{ij}=g(a)S^*_{ij}$. In a matrix form the system of
Eqs. (\ref{n2}) can be written as follows:

\begin{equation}\label{n4}
\begin{split}
& \left(
    \begin{array}{cccc}
   1-A_1 K_{11} & 0 & \cdots & 0 \\
    -A_1 K_{21} & 1-A_2 K_{22} & \cdots & 0 \\
    \vdots &  & \ddots &  \\
    - A_1 K_{n1} & - A_2 K_{n2} & \cdots & 1- A_n K_{nn} \\
  \end{array}
\right)
 \left(
                       \begin{array}{c}
                         j^*_1 \\
                         j^*_2 \\
                         \vdots \\
                         j^*_n
                       \end{array}
                     \right)=\\
                     & = \left(
                     \begin{array}{c}
                         1 \\
                         1 \\
                         \vdots \\
                         1 \\
                       \end{array}
                       \right).
 \end{split}
\end{equation}

Using the unequally spaced mesh with an increment $h_{i}=\tau_{i} -
\tau_{i-1},\;\; i=2,...,n$ and applying the trapezoidal
integration rule Eqs. (\ref {n2}) yield the following recurrence
equations:
\begin{equation}\label{n5}
\begin{split}
 & j^*_1 = 1 \\
 & j^*_2 = \D \frac{1+\displaystyle \frac{h_2}{2} K_{21}j^*_1}{1-\D \frac{h_2}{2} K_{22}}\\
 & j^*_i = \D \frac{1+\D\frac{h_2}{2}K_{i1}j^*_1+\D \sum_{j=2}^{i-1}\left( \frac{\tau_{j+1}-\tau_{j-1}}{2}\right)K_{ij}j^*_j}{1-\D
 \frac{h_i}{2}K_{ii}}\;\; (i=3,...,n)
 \end{split}
\end{equation}
Equations (\ref{n5}) are valid in the case when $\displaystyle
h_i\neq\frac{2}{K_{ii}}$.

In the numerical calculations we spaced the mesh points adaptively
using the following formula:
\begin{equation}\label{mshp}
    \tau_i = \tau_1 + \left(\tau_N - \tau_1\right)\left[1 - \cos \left(\frac{\pi}{2}
\frac{i-1}{N-1}\right)\right],\;\;\;\; (i=1,2,...,N) \; .
\end{equation}
In \re{mshp} $N$ is the chosen number of mesh points, $\tau_1$ and $\tau_N$ are the locations of  left and right boundaries of time interval, respectively.

\section{Results and discussion}

Using the suggested model the calculations were performed for sulfur dioxide ($\mathrm {SO}_2$), dinitrogen trioxide (N$_{2}$O$_{3}$) and chlorine (Cl$_{2}$) absorption by water aerosol particles. In order to validate our model we compared the results obtained using the suggested model with the results obtained in our previous study for large droplets ($Kn\ll1$)\citep[see e.g.,][]{Elperin2008}. The calculations were performed for the  $\mathrm {SO}_2$ absorption by a non--evaporating water droplet of 10 $\mu$m in radius. The concentration of sulfur dioxide in  ambient air was assumed to be 0.01 ppm. The calculations showed that the time of the complete saturation of droplet by sulfur dioxide estimated using the suggested model is $\approx$ 0.08 s, while the time of complete saturation of droplet by sulfur dioxide estimated using our previous model is $\approx$ 0.1 s. These calculations demonstrate that the results obtained by both models are in fairly good agreement.

\begin{figure}[h!]
 \begin{center}
  \includegraphics[width= 8 cm]{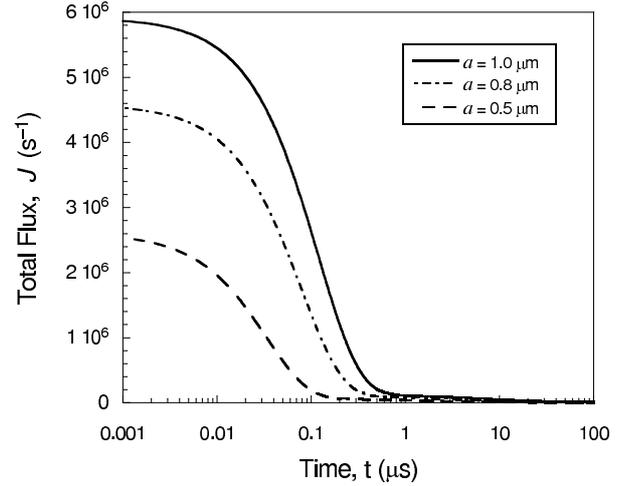}
  \caption[Figure 2]{Total molecular flux of sulfur dioxide as a function of time (droplet radii $a =$ 1.0, 0.8 and 0.5 $\mu$m)} \label{fig2}
  \end {center}
\end{figure}

\begin{figure}[h!]
 \begin{center}
  \includegraphics[width= 8 cm]{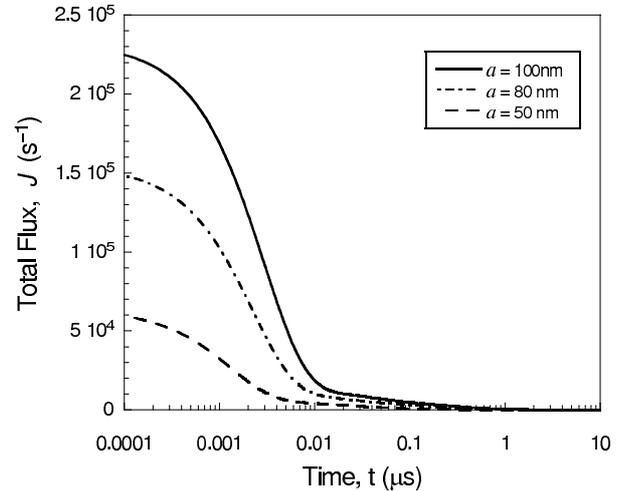}
  \caption[Figure 3]{Total molecular flux of sulfur dioxide as a function of time (droplet radii $a =$ 100.0, 80.0 and 50.0 nm)} \label{fig3}
  \end {center}
\end{figure}

The results of calculation of the total mass flux of sulfur dioxide as a function of time are shown in Figs. \ref{fig2} and \ref{fig3}. The calculations were performed for various radii of water aerosol particle (from 0.5 $\mu\textnormal m$ to 1.0 $\mu\textnormal m$, $0.07 \lesssim Kn \lesssim 0.14$, see Fig. \ref{fig2} and from 50 nm to 100 nm, 0.7 $\lesssim Kn \lesssim 1.42$, see Fig. \ref{fig3}).

As can be seen from these plots for small and moderate size droplets the flux of absorbate decreases rapidly at the initial stage of gas absorption and approaches asymptotically zero at the final stage of the process. The vanishing flux of the absorbate implies the stage of saturation of droplet by gas.

\begin{figure}[h!]
 \begin{center}
  \includegraphics[width= 8 cm]{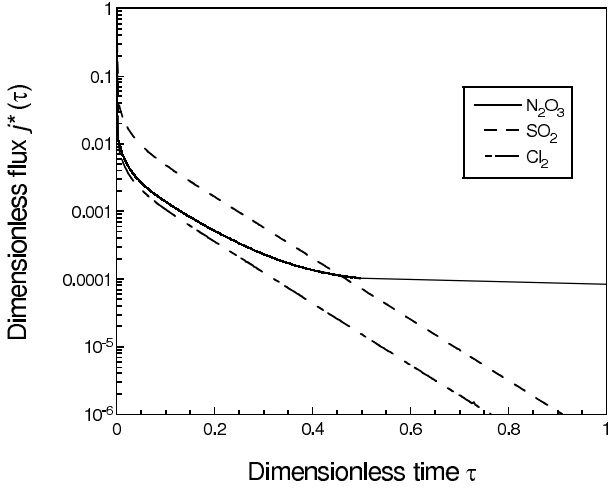}
  \caption[ Figure 4]{Dimensionless molecular flux density $j^*(\tau)$ as a function of dimensionless time $\tau$ (radius of the droplet $R$ = 100 nm).} \label{fig4}
  \end {center}
\end{figure}

The results of calculation of the dimensionless flux $\displaystyle j^\ast (\tau)= j(t)4\pi a^2/\alpha(a)n_\infty$ as a function of dimensionless time $\tau = D_{\sub L}t/a^2$ for different gases such as sulfur dioxide, chlorine and dinitrogen trioxide are shown in Fig. \ref{fig4}. Larger values of mass flux at the later stages of gas absorption for N$_{2}$O$_{3}$ in comparison with $\mathrm {SO}_2$ and $\mathrm {Cl}_2$ absorption can be explained by large values of the constant of chemical reaction for N$_{2}$O$_{3}$ gas absorption in water ($\lambda_{\mathrm N_{2}\mathrm O_{3}} = 1.2\cdot 10^4$ s$^{-1}$, $\lambda_{\mathrm {SO}_{2}} = 10^{-3}$ s$^{-1}$, and $\lambda_{\mathrm {Cl}_{2}} = 13.3$ s$^{-1}$).

Intensive depletion of the dissolved gaseous N$_{2}$O$_{3}$ in a water droplet due to chemical reaction leads to the
decrease of N$_{2}$O$_{3}$ concentration in the bulk of a water droplet and to increase of the concentration gradient at the interface in a liquid phase. These both factors increase mass transfer coefficient in a droplet and increase the driving force of mass transfer in liquid.

\begin{figure}[h!]
 \begin{center}
  \includegraphics[width= 8 cm]{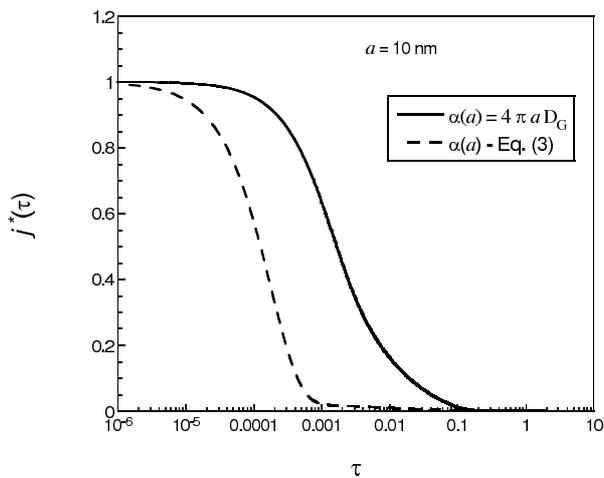}
  \caption[Figure 5]{Dimensionless molecular flux density $j^*(\tau)$ as a function of dimensionless time $\tau$ (plane line -- $\alpha(a) = 4 \pi a D_G $, dashed line -- $\alpha(a)$ calculated using Eq. (\ref {4})).} \label{fig5}
  \end {center}
\end{figure}

As it was mentioned above in the case of large droplets ($Kn \ll 1$) the capture efficiency (see \rr{4}{5}) can be expressed by $\alpha (a) = 4\pi a D_G$. Consequently the flux of soluble gas in a gaseous phase is expressed by equation similar to  Maxwell's equation
 (see Eq. (\ref{maxwell})). In Fig. \ref{fig5} we showed the results of calculation of the dimensionless flux $j^\ast(\tau)$ of sulfur dioxide as a function of the dimensionless time $\tau$. It was assumed that concentration of sulfur dioxide in a gaseous phase is equal to 1 ppb, temperature in a gaseous phase 298 K and radius of water nanoparticle is equal to 10 nm. The dashed line presents the results of calculation when the capture efficiency was calculated using \re{5}. In our calculations we assumed that the accommodation coefficient $S_p = 1$.

The solid line present the results of calculation without using the kinetic approach, and thereby the capture efficiency was assumed to be equal $\alpha (a) = 4\pi a D_G$. As can be seen from these plots neglecting kinetic effects in the case of gas absorption by nanoaerosols can lead to the essential overestimation of mass flux.

\begin{figure}[h!]
 \begin{center}
  \includegraphics[width= 8 cm]{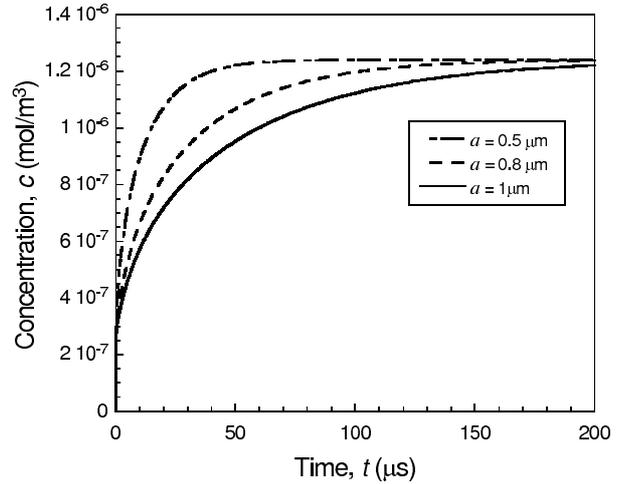}
  \caption[Figure 6]{Concentration of the dissolved $\mathrm {SO_2}$ in the bulk of a droplet as a function of time (radii of a droplet 0.5, 0.8 and 1.0 $\mu\mathrm {m}$).} \label{fig6}
  \end {center}
\end{figure}

\begin{figure}[h!]
 \begin{center}
  \includegraphics[width= 8 cm]{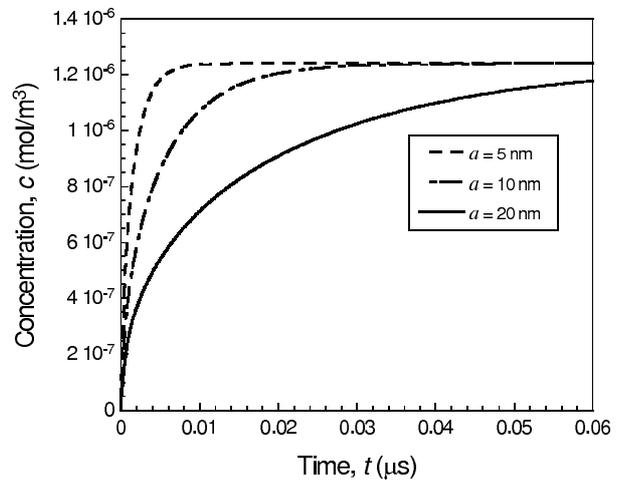}
  \caption[Figure 7]{Concentration of the dissolved $\mathrm {SO_2}$ in the bulk of a droplet as a function of time (radii of a droplet 5.0, 10.0 and 20.0 nm).}
  \label{fig7}
  \end {center}
\end{figure}

Dependence of the average concentration of soluble sulfur dioxide in a droplet vs. time is shown in Figs. \ref{fig6} and \ref{fig7}.

\begin{figure}[h!]
 \begin{center}
  \includegraphics[width= 8 cm]{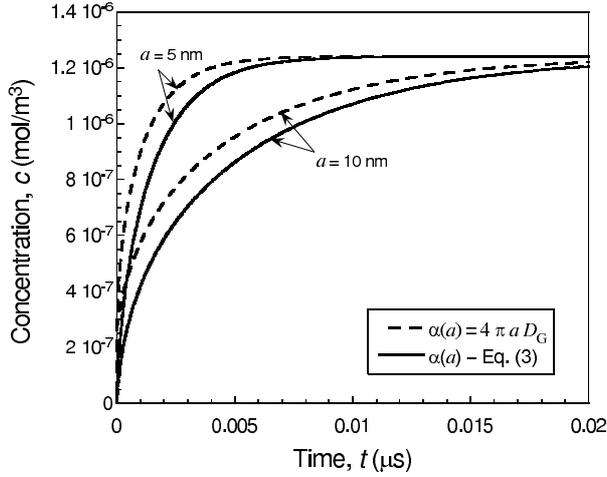}
  \caption[Figure 8]{Effect of Knudsen layer on temporal evolution of concentration of the dissolved SO$_2$ for droplets with the radii 5.0 and 10.0 nm.}
  \label{fig8}
  \end {center}
\end{figure}

Calculations were performed for water droplets with the radii 0.5, 0.8 and 1 $\mu$m (Fig. \ref{fig6}) and for water droplets of radii 5, 10 and 20 nm (Fig. \ref{fig7}). In these calculations we employed the kinetic approach by using \re{5} for the capture efficiency $\alpha (a)$ with $S_p = 1$. In Fig. \ref{fig8} the dependence of average concentration of SO$_{2}$ in a droplet vs. time was calculated using the kinetic approach (solid lines) and neglecting kinetic effects (dashed lines).
Calculations were performed for the droplets with the radii 10 and 5 nm. As can be seen from these plots neglecting kinetic effects leads to the significant overestimation of concentration of the dissolved gas in a droplet during all the period of gas absorption. Clearly, when the duration of gas absorption $t\rightarrow \infty$, both approaches yield the same result for the magnitude of the dissolved gas concentration.

\section{Conclusions}

In this study we developed a model for absorption of soluble trace gases by nanoaerosols taking into account dissociation reaction of the first order in a liquid phase. In the case when radius of the particle is comparable with the mean free path transport of reactant molecules cannot be described by Fickian diffusion. However, application of the flux-matching theory allowed using transient diffusion equation with the kinetic boundary conditions for the description of gas absorption by nanoaerosols. Transient diffusion equation was solved analytically by the method of separation of variables. We derived linear integral equation of Volterra type for the transient mass flux to a liquid droplet. Integral equation was solved numerically by the method based on the approximation of the integral using the quadrature formula with unequally spaced mesh.

The comparison of the suggested model with our earlier model developed for gas absorption by large droplets ($Kn\ll1$) (see, Elperin eta al., 2008) showed that both models yield the same time of complete saturation of a large droplet by the soluble gas.

Using the suggested model we studied absorption of sulfur dioxide ($\mathrm {SO}_2$), dinitrogen trioxide (N$_{2}$O$_{3}$) and chlorine (Cl$_2$) by water nanoaerosol. It is showed that enhanced depletion of the dissolved N$_{2}$O$_{3}$ gas in a water droplet due to chemical reaction leads to the decrease of N$_{2}$O$_{3}$ concentration in the bulk of a water droplet and to the increase of the concentration gradient at the gas--liquid interface. Consequently, the flux of dinitrogen trioxide into a droplet is higher than the fluxes of sulfur dioxide and chlorine at later stages of gas absorption.

It was demonstrated that neglecting kinetic effects leads to the significant overestimation of the soluble gas flux into a droplet during all the
period of gas absorption.

\appendix
\section{Derivation of the integral equation of Volterra type for the molecular flux density}

Here we give the details of derivation of \re {j}. Let us first remove $j$ from the boundary condition (\ref{3a}). To this end let us introduce the new unknown function $C(r,t)$:
\be
n_L=C(r,t)-\frac{j(t)r}{D_L}.
\ee{4a}
Substituting \re{4a} into \re{1a}, and taking into account that in spherical coordinates the Laplacian $\Delta r=2/r$ we obtain the following equation for $C(r,t)$:
\be
\pa{C}{t}=D_L\Delta C+ \frac{r j_t}{D_L}-\frac{2 j}r-\lambda C+\frac{\lambda j r}{D_L}\:,
\ee{5a}
with the following boundary conditions
\be
\left.\pa{C}{r}\right|_{r=a}=0
\ee{5aa}
and
\begin{equation}\label{5bb}
    \left.\frac{\partial C}{\partial r}\right|_{r = 0}=\left.\frac{j}{D_L}\right|_{r = 0}\;.
\end{equation}
Substitution
\be
C(r,t)=\frac{\chi (r,t)}r
\ee{6aa}
reduces \re{5a} to
\be
\pa{\chi}{t}=D_L\pa{^2\chi}{r^2}-\lambda\chi-2j+\frac1{D_L}(r^2 j_t + \lambda jr^2)\:.
\ee{7aa}
The boundary conditions to Eq. (\ref{7aa}) read:
\be
\left.\pa{\chi}{r}\right|_{r=a}-\left.\frac \chi r\right|_{r=a}=0
\ee{7aaa}
and
\begin{equation}\label{7bbb}
    \left.\chi \right|_{r=0}=0.
\end{equation}
Let us introduce the eigenfunctions
\be
\od{^2\phi_n}{r^2} =-\kappa^2\phi_n \:,
\ee{8aa}
where the boundary conditions to the equation (\ref{8aa}) are the same as for $\chi$, i.e., given by Eqs. (\ref{7aaa}) -- (\ref{7bbb}):
\be
\left.\pa{\phi_n}{r}\right|_{r=a}-\left.\frac {\phi_n} {r}\right|_{r=a}=0\; , \qquad \left.\phi_n \right|_{r=0}=0.
\ee{9b}
Then the solution of the \re{8aa} reads:
\be
\phi_n=u_n\sin(\mu_n\frac{r}{a})\:,
\ee{10aa}
where $\mu = \kappa a$ is the infinite set of the roots of the characteristic equation:
\be
\mu =\tan(\mu)\:.
\ee{11aa}
The roots $\mu_n$ of the \re{11aa} can be calculated numerically and are as follows $\mu_1 = 4.4934, \mu_2 = 7.7253, \mu_3 = 10.9041, \mu_4 = 14.0662, \mu = 17.2208$ etc.
The orthogonality condition for eigenfunctions reads:
\be
\inl_0^a\phi_n\phi_{m}dr=\delta_{n m}
\ee{12aa}
where $\delta_{n m}$ is the Kronecker delta.
Equation (\ref{12aa}) allows to determine the normalization constant $u_n$:
\be
u_n^2=\left(\inl_0^a \sin^2(\mu_n \frac{r}{a})dr\right)^{-1}=\frac2{a\sin^2\mu_n}\:.
\ee{13aa}
The eigenvalue $\mu_n = 0$ (which is also the solution of \re{11aa}) and the respective eigenfunction require a special consideration. Solution of \re{8aa} for $\mu_n = 0$ reads:
\be
\phi_0(r)=u_0r\:,
\ee{14aa}
where the normalization constant $u_0$ is determined from \re{12aa}:
\be
u_0=\sqrt{\frac3{a^3}}\:.
\ee{15aa}
Let us now look for the solution to \re{7aa} in the following form:
\be
\chi(r,t)=2 \Psi_0(t)\phi_0+\sml_{n > 0} \Psi_n(t)\phi_n(r)\:,
\ee{13a}
where the coefficient $2$ appears due to double degeneration of the eigenvalue $\mu_n =0$.
Equation for $\Psi_n$ reads:
\be
\od{\Psi_n}{t}=-\sigma_n \Psi_n + \frac1{D_L} (j_t+\lambda j)b_n-2ja_n\:,
\ee{14a}
where

\begin{equation}\label{15a}
  \begin{split}
    \sigma_n= \lambda + & D_L\left(\frac{\mu_n}{a}\right)^2, \qquad a_n =u_n\inl_0^a\sin\left(\mu_n \frac{r}{a}\right)dr,\\
    & b_n=u_n\inl_0^ar^2\sin\left(\mu_n \frac{r}{a}\right)dr
  \end{split}
\end{equation}

and
\be
\begin{split}
\sigma_0=\lambda,& \qquad a_0 =u_0\inl_0^ardr=\frac12\sqrt{3a},\\
& b_0=u_0\inl_0^a r^3dr=\frac14\sqrt{3{a^5}}.
\end{split}
\ee{15aa}
 For $n > 0$ we obtain:
\be
\Psi_n(t)=\frac{b_nj(t)}{D_L}-\left(\left(\frac{\mu_n}{a}\right)^2 b_n+ 2a_n\right)
\inl_0^tj(t')e^{-\sigma_n (t-t')}dt'
\ee{16a}
and
\be
\Psi_0(t)=\frac{b_0 j(t)}{D_L}- 2a_0
\inl_0^tj(t')e^{-\lambda (t-t')}dt'
\ee{16aa}
for $n=0$. The number density of molecules $A$ inside the particle near the surface reads:
\be
n_{-}(a,t)=\frac{\chi(a,t)}a-\frac{j(t)a}{D_{\sub L}}\:.
\ee{17a}
Now let us determine $\chi$ from \re{13a}. Noting that
\be
\sml_{n\ge0} b_n\phi_n(a) =\frac{a^2}{4}
\ee{17aa}
and
\be
\left(\left(\frac{\mu_n}{a}\right)^2b_n+ 2a_n\right)\phi_n(a)=2\quad n>0\:,
\ee{17ab}
\be
\left(\left(\frac{\mu_n}{a}\right)^2b_n+ 2a_n\right)\phi_n(a)=3\quad n=0\:,
\ee{17ac}
\re{13a} and \re{17a} yield:
\be
n_-(a)=-2\sml_{n > 0} \inl_0^te^{-\sigma_n(t- t')}j(t')dt'-3\inl_0^te^{-\lambda(t- t')}j(t')dt'.
\ee{17ad}
According to the Henry's law $n_+={\cal H}n_-$. Using the equation for the flux density
\be
j(t)=\frac1{4\pi a^2}\alpha(a)(n_\infty-n_+)
\ee{18a}
we arrive at the following integral equation of Volterra type for $j(t)$:
\be
j(t)=\frac{\alpha(a)}{4\pi a^2}\left[n_\infty-{\cal H}\inl_0^tS(t-t')j(t')dt'\right],
\ee{19a}
where the kernel $S(t-t')$ is given by the following formula:
\be
S(\xi)=2\sml_{n >\;0} e^{-[D_L(\mu_n / a)^2+\lambda]\xi}+3 e^{-\lambda\xi}\:.
\ee{20a}

\bibliographystyle{elsarticle-harv}
%%\bibliography{<your-bib-database>}




%% \bibitem must have one of the following forms:
%%   \bibitem[Jones et al.(1990)]{key}...
%%   \bibitem[Jones et al.(1990)Jones, Baker, and Williams]{key}...
%%   \bibitem[Jones et al., 1990]{key}...
%%   \bibitem[\protect\citeauthoryear{Jones, Baker, and Williams}{Jones
%%       et al.}{1990}]{key}...
%%   \bibitem[\protect\citeauthoryear{Jones et al.}{1990}]{key}...
%%   \bibitem[\protect\astroncite{Jones et al.}{1990}]{key}...
%%   \bibitem[\protect\citename{Jones et al., }1990]{key}...
%%   \harvarditem[Jones et al.]{Jones, Baker, and Williams}{1990}{key}...
%%


% \end{thebibliography}

%\end{linenumbers}

\end{document}